\documentclass[12pt]{iopart}

\usepackage{cite}
\usepackage{graphicx}

\begin{document}

\title{Harmonic oscillator in a one--dimensional box}

\author{Paolo Amore\dag \ and Francisco M Fern\'andez
\footnote[2]{Corresponding author}}

\address{\dag\ Facultad de Ciencias, Universidad de Colima, Bernal D\'iaz del
Castillo 340, Colima, Colima, Mexico}\ead{paolo@ucol.mx}

\address{\ddag\ INIFTA (UNLP, CCT La Plata-CONICET), Divisi\'on Qu\'imica Te\'orica,
Blvd. 113 S/N,  Sucursal 4, Casilla de Correo 16,
1900 La Plata, Argentina}\ead{fernande@quimica.unlp.edu.ar}

\maketitle

\begin{abstract}
We study a harmonic molecule confined to a one--dimensional box with
impenetrable walls. We explicitly consider the symmetry of the problem for
the cases of different and equal masses. We propose suitable variational
functions and compare the approximate energies given by the variation method
and perturbation theory with accurate numerical ones for a wide range of
values of the box length. We analyze the limits of small and large box size.
\end{abstract}

\section{Introduction\label{sec:intro}}

During the last decades there has been great interest in the model of a
harmonic oscillator confined to boxes of different shapes and sizes\cite
{AK40,A41,A42,C43,AK45,D52,BS55,D66,V68,V73,CF76,AM77,R78,ALZ80,AILZ81,
BR81,FC81a,FC81b,AGZL83,CM83,ML83,A97,V98}. Such model has been suitable for
the study of several physical problems ranging from dynamical friction in
star clusters\cite{C43} to magnetic properties of solids\cite{D52} and
impurities in quantum dots\cite{V98}.

One of the most widely studied model is given by a particle confined to a
box with impenetrable walls at $-L/2$ and $L/2$ bound by a linear force that
produces a parabolic potential--energy function $V(x)=k(x-x_{0})^{2}/2$,
where $|x_{0}|<L/2$. When $x_{0}=0$ the problem is symmetric and the
eigenfunctions are either even or odd; such symmetry is broken when $%
x_{0}\neq 0$. Although interesting in itself, this model is rather
artificial because the cause of the force is not specified. It may, for
example, arise from an infinitely heavy particle clamped at $x_{0}$. In such
a case we think that it is more interesting to consider that the other
particle also moves within the box.

The purpose of this paper is the discussion of the model of two particles
confined to a one--dimensional box with impenetrable walls. For simplicity
we assume that the force between them is linear. In Sec.~\ref{sec:model} we
introduce the model and discuss some of its general mathematical properties.
In Sec.~\ref{sec:small_L} we discuss the solutions of the Schr\"{o}dinger
equation for small box lengths by means of perturbation theory. In Sec.~\ref
{sec:large_L} we consider the regime of large boxes and propose suitable
variational functions. In Sec.~\ref{sec:results} we compare the approximate
energies provided by perturbation theory and the variational method with
accurate numerical ones. Finally, in Sec.~\ref{sec:conclusions}~we summarize
the main results and draw additional conclusions.

\section{The Model\label{sec:model}}

As mentioned above, we are interested in a system of two particles of masses
$m_{1}$ and $m_{2}$ confined to a one--dimensional box with impenetrable
walls located at $x=-L/2$ and $x=L/2$. If we assume a linear force between
the particles then the Hamiltonian operator reads
\begin{equation}
\hat{H}=-\frac{\hbar ^{2}}{2}\left( \frac{1}{m_{1}}\frac{\partial ^{2}}{%
\partial x_{1}^{2}}+\frac{1}{m_{2}}\frac{\partial ^{2}}{\partial x_{2}^{2}}%
\right) +\frac{k}{2}(x_{1}-x_{2})^{2}  \label{eq:H}
\end{equation}
and the boundary conditions are $\psi =0$ when $x_{i}=\pm L/2$. It is
convenient to convert it to a dimensionless form by means of the variable
transformation $q_{i}=x_{i}/L$ that leads to:
\begin{equation}
\hat{H}_{d}=\frac{m_{1}L^{2}}{\hbar ^{2}}\hat{H}=-\frac{1}{2}\left( \frac{%
\partial ^{2}}{\partial q_{1}^{2}}+\beta \frac{\partial ^{2}}{\partial
q_{2}^{2}}\right) +\frac{\lambda }{2}\left( q_{1}-q_{2}\right) ^{2}
\label{eq:Hd}
\end{equation}
where $\beta =m_{1}/m_{2}$, $\lambda =km_{1}L^{4}/\hbar ^{2}$ and the
boundary conditions become $\psi =0$ if $q_{i}=\pm 1/2$. Without loss of
generality we assume that $0<\beta \leq 1$.

The free problem ($-\infty <x_{i}<\infty $) is separable in terms of
relative and center--of--mass variables
\begin{eqnarray}
x &=&x_{1}-x_{2}  \nonumber \\
X &=&\frac{1}{M}\left( m_{1}x_{1}+m_{2}x_{2}\right) ,\;M=m_{1}+m_{2}
\label{eq:CM}
\end{eqnarray}
respectively, that lead to
\begin{equation}
\hat{H}=-\frac{\hbar ^{2}}{2}\left( \frac{1}{M}\frac{\partial ^{2}}{\partial
X^{2}}+\frac{1}{m}\frac{\partial ^{2}}{\partial x^{2}}\right) +V(x),\;m=%
\frac{m_{1}m_{2}}{M}.  \label{eq:H_CM}
\end{equation}
In this case we can factor the eigenfunctions as
\begin{eqnarray}
\psi _{Kv}(x_{1},x_{2}) &=&e^{iKX}\phi _{v}(x)  \nonumber \\
-\infty &<&K<\infty ,\;v=0,1,\ldots  \label{eq:Psi_CM}
\end{eqnarray}
where $\phi _{v}(x)$ are the well--known eigenfunctions of the harmonic
oscillator and the eigenvalues read
\begin{equation}
E_{Kv}=\frac{\hbar ^{2}K^{2}}{2M}+\hbar \sqrt{\frac{k}{m}}\left( v+\frac{1}{2%
}\right) .  \label{eq:E_Kv}
\end{equation}
However, because of the boundary conditions, any eigenfunction is of the
form $\psi (x_{1},x_{2})=(L^{2}/4-x_{1}^{2})(L^{2}/4-x_{2}^{2})\Phi
(x_{1},x_{2})$, where $\Phi (x_{1},x_{2})$ does not vanish at the walls. We
clearly appreciate that the separation just outlined is not possible in the
confined model.

When $\beta <1$ the transformations that leave the Hamiltonian operator
(including boundary conditions) invariant are: identity $\hat{E}%
:(q_{1},q_{2})\rightarrow (q_{1},q_{2})$ and inversion $\hat{\imath}%
:(q_{1},q_{2})\rightarrow (-q_{1},-q_{2})$. Therefore, the eigenfunctions of
$\hat{H}_{d}$ are basis for the irreducible representations $A_{g}$ and $%
A_{u}$ of the point group $S_{2}$\cite{T64} (also called $C_{i}$ by other
authors).

On the other hand, when $\beta =1$ (equal masses) the problem exhibits the
highest possible symmetry. The transformations that leave the Hamiltonian
operator (including boundary conditions) invariant are: identity $\hat{E}%
:(q_{1},q_{2})\rightarrow (q_{1},q_{2})$, rotation by $\pi $ $%
C_{2}:(q_{1},q_{2})\rightarrow (q_{2},q_{1})$, inversion $\hat{\imath}%
:(q_{1},q_{2})\rightarrow (-q_{1},-q_{2})$, and reflection in a plane
perpendicular to the rotation axis $\sigma _{h}:(q_{1},q_{2})\rightarrow
(-q_{2},-q_{1})$. In this case the states are basis functions for the
irreducible representations $A_{g}$, $A_{u}$, $B_{g}$, and $B_{u}$ of the
point group $C_{2h}$\cite{T64}.

\section{Small box\label{sec:small_L}}

When $\lambda \ll 1$ we can apply perturbation theory choosing the
unperturbed or reference Hamiltonian operator to be $\hat{H}_{d}^{0}=\hat{H}%
_{d}(\lambda =0)$. Its eigenfunctions and eigenvalues are given by
\begin{eqnarray}
\varphi _{n_{1},n_{2}}^{(0)}(q_{1},q_{2}) &=&\left\{
\begin{array}{ll}
2\cos [(2i-1)\pi q_{1}]\cos [(2j-1)\pi q_{2}] & A_{g} \\
2\sin (2i\pi q_{1})\sin (2j\pi q_{2}) & A_{g} \\
2\cos [(2i-1)\pi q_{1}]\sin (2j\pi q_{2}) & A_{u} \\
2\sin (2i\pi q_{1})\cos [(2j-1)\pi q_{2}] & A_{u}
\end{array}
\right. ,\;i,j=1,2,\ldots  \nonumber \\
\epsilon _{n_{1},n_{2}}^{(0)} &=&\frac{\pi ^{2}}{2}\left( n_{1}^{2}+\beta
n_{2}^{2}\right) ,\;n_{1},\,n_{2}=1,2,\ldots .  \label{eq:zeroth-order}
\end{eqnarray}
There is no degeneracy when $\beta <1$, except for the accidental one that
takes place for particular values of $\beta $ which we will not discuss in
this paper. The energies corrected to first order read
\begin{equation}
\epsilon _{n_{1},n_{2}}^{[1]}=\frac{\pi ^{2}}{2}\left( n_{1}^{2}+\beta
n_{2}^{2}\right) +\lambda \frac{\pi ^{2}n_{1}^{2}n_{2}^{2}-3\left(
n_{1}^{2}+n_{2}^{2}\right) }{12\pi ^{2}n_{1}^{2}n_{2}^{2}}.
\label{eq:e_PT_nond}
\end{equation}

When $\beta =1$ the zeroth--order states $\varphi _{n_{1},n_{2}}^{(0)}$ and $%
\varphi _{n_{2},n_{1}}^{(0)}$ ($n_{1}\neq n_{2}$) are degenerate but it is
not necessary to resort to perturbation theory for degenerate states in
order to obtain the first--order energies. We simply take into account that
the eigenfunctions of $\hat{H}_{d}^{0}$ adapted to the symmetry of the
problem are
\begin{equation}
\varphi _{n_{1},n_{2}}^{(0)}(q_{1},q_{2})=\left\{
\begin{array}{ll}
\sqrt{2-\delta _{ij}}\left\{ \cos [(2i-1)\pi q_{1}]\cos [(2j-1)\pi
q_{2}]\right. &  \\
\left. +\cos [(2j-1)\pi q_{1}]\cos [(2i-1)\pi q_{2}]\right\} & A_{g} \\
&  \\
\sqrt{2}\left\{ \cos [(2i-1)\pi q_{1}]\cos [(2j-1)\pi q_{2}]\right. &  \\
\left. -\cos [(2j-1)\pi q_{1}]\cos [(2i-1)\pi q_{2}]\right\} & B_{g} \\
&  \\
\sqrt{2}\left\{ \cos [(2i-1)\pi q_{1}]\sin [2j\pi q_{2}]+\sin [2j\pi
q_{1}]\cos [(2i-1)\pi q_{2}]\right\} & A_{u} \\
&  \\
\sqrt{2}\left\{ \cos [(2i-1)\pi q_{1}]\sin [2j\pi q_{2}]-\sin [2j\pi
q_{1}]\cos [(2i-1)\pi q_{2}]\right\} & B_{u} \\
&  \\
\sqrt{2-\delta _{ij}}\left\{ \sin [2i\pi q_{1}]\sin [2j\pi q_{2}]+\sin
[2j\pi q_{1}]\sin [2i\pi q_{2}]\right\} & A_{g} \\
&  \\
\sqrt{2}\left\{ \sin [2i\pi q_{1}]\sin [2j\pi q_{2}]-\sin [2j\pi q_{1}]\sin
[2i\pi q_{2}]\right\} & B_{g}
\end{array}
\right.  \label{eq:fi0_sym}
\end{equation}
where $i,j=1,2,\ldots $. They give us the energies corrected to first order
as $\epsilon _{n}^{[1]}(S)=\left\langle \varphi _{ij}^{(0)}(S)\right| \hat{H}%
_{d}\left| \varphi _{ij}^{(0)}(S)\right\rangle $ where $S$ denotes the
irreducible representation. Since some of these analytical expressions are
rather cumbersome for arbitrary quantum numbers, we simply show the first
sixth energy levels for future reference:
\begin{eqnarray}
\epsilon _{1}^{[1]}(A_{g}) &=&\pi ^{2}+\frac{\lambda (\pi ^{2}-6)}{12\pi ^{2}%
}  \nonumber \\
\epsilon _{1}^{[1]}(A_{u}) &=&\frac{5\pi ^{2}}{2}+\frac{\lambda (108\pi
^{4}-405\pi ^{2}-4096)}{1296\pi ^{4}}  \nonumber \\
\epsilon _{1}^{[1]}(B_{u}) &=&\frac{5\pi ^{2}}{2}+\frac{\lambda (108\pi
^{4}-405\pi ^{2}+4096)}{1296\pi ^{4}}  \nonumber \\
\epsilon _{2}^{[1]}(A_{g}) &=&4\pi ^{2}+\frac{\lambda (2\pi ^{2}-3)}{24\pi
^{2}}  \nonumber \\
\epsilon _{3}^{[1]}(A_{g}) &=&\epsilon _{1}^{[1]}(B_{g})=5\pi ^{2}+\frac{%
\lambda (3\pi ^{2}-10)}{36\pi ^{2}}.  \label{eq:e_PT_deg}
\end{eqnarray}
The degeneracy of the approximate energies denoted $\epsilon
_{3}^{[1]}(A_{g})$ and $\epsilon _{1}^{[1]}(B_{g})$ is broken at higher
perturbation orders as shown by the numerical results in Sec.~\ref
{sec:results}.

\section{Large box\label{sec:large_L}}

When $L\rightarrow \infty $ the energy eigenvalues tend to those of the free
system (\ref{eq:E_Kv}). More precisely, we expect that the states with
finite quantum numbers $n_{1}$, $n_{2}$ at $L=0$ correlate with those with $%
K=0$ when $L\rightarrow \infty $:
\begin{equation}
\lim_{\lambda \rightarrow \infty }\frac{\epsilon (\beta ,\lambda )}{\sqrt{%
\lambda }}=\sqrt{1+\beta }\left( v+\frac{1}{2}\right) ,\;v=0,1,\ldots .
\label{eq:e_lim}
\end{equation}
Besides, we should take into account that the symmetry of a given state is
conserved as $L$ increases from $0$ to $\infty $.

When $\beta <1$ we expect that the states approach
\begin{equation}
\psi _{Kv}(x,X)=\left\{
\begin{array}{ll}
\cos (KX)\phi _{2v}(x) & A_{g} \\
\sin (KX)\phi _{2v+1}(x) & A_{g} \\
\sin (KX)\phi _{2v}(x) & A_{u} \\
\cos (KX)\phi _{2v+1}(x) & A_{u}
\end{array}
\right.  \label{eq:Psi_lim_bet<1}
\end{equation}
as $L\rightarrow \infty $.

For the more symmetric case $\beta =1$ the states should be
\begin{equation}
\psi _{Kv}(x,X)=\left\{
\begin{array}{ll}
\cos (KX)\phi _{2v}(x) & A_{g} \\
\sin (KX)\phi _{2v}(x) & A_{u} \\
\cos (KX)\phi _{2v+1}(x) & B_{u} \\
\sin (KX)\phi _{2v+1}(x) & B_{g}
\end{array}
\right. .  \label{eq:Psi_lim_bet=1}
\end{equation}
Obviously, the perturbation expressions (\ref{eq:e_PT_nond}) or (\ref
{eq:e_PT_deg}) are unsuitable for this analysis and we have to resort to
other approaches.

In order to obtain accurate eigenvalues and eigenfunctions for the present
model we may resort to the Rayleigh--Ritz variational method and the basis
set of eigenfunctions of $\hat{H}_{d}^{0}$ given in equations (\ref
{eq:zeroth-order}) and (\ref{eq:fi0_sym}). Alternatively, we can also make
use of the collocation method with the so--called little sinc functions
(LSF) that proved useful for the treatment of coupled anharmonic oscillators%
\cite{AF09}. In this paper we choose the latter approach.

Another way of obtaining approximate eigenvalues and eigenfunctions is
provided by a straightforward variational method proposed some time ago\cite
{AFC84}. The trial functions suitable for the present model are of the form
\begin{equation}
\varphi (q_{1},q_{2})=\left( \frac{1}{4}-q_{1}^{2}\right) \left( \frac{1}{4}%
-q_{2}^{2}\right) f(\mathbf{c},q_{1},q_{2})e^{-a(q_{1}-q_{2})^{2}}
\label{eq:fi_var}
\end{equation}
where $\mathbf{c}=\{c_{1},c_{2},\ldots ,c_{N}\}$ are linear variational
parameters, which would give rise to the well known Rayleigh--Ritz secular
equations, and $a$ is a nonlinear one. Even the simplest and crudest
variational functions provide reasonable results for all values of $\lambda $
as shown in Sec.~\ref{sec:results}.

The simplest trial function for the ground state of the model with $\beta <1$
is
\begin{equation}
\varphi (q_{1},q_{2})=\left( \frac{1}{4}-q_{1}^{2}\right) \left( \frac{1}{4}%
-q_{2}^{2}\right) e^{-a(q_{1}-q_{2})^{2}}.  \label{eq:fi_var_bet<>1}
\end{equation}
Notice that this function is basis for the irreducible representation $A_{g}$%
. We calculate $w(a,\lambda )=\left\langle \varphi \right| \hat{H}_{d}\left|
\varphi \right\rangle /\left\langle \varphi \right. \left| \varphi
\right\rangle $ and obtain $\lambda (a)$ from the variational condition $%
\partial w/\partial a=0$ so that $[w(a,\lambda (a)),\lambda (a)]$ is a
suitable parametric representation of the approximate energy. In this way we
avoid the tedious numerical calculation of $a$ for each given value of $%
\lambda $ and obtain an analytical parametric expression for the energy that
we do not show here because it is rather cumbersome. We just mention that
the parametric expression is valid for $a>a_{0}$ where $a_{0}$ is the
greatest positive root of $\lambda (a)=0$.

When $\beta =1$ we choose the following trial functions for the lowest
states of each symmetry type
\begin{eqnarray}
\varphi _{A_{g}}(q_{1},q_{2}) &=&\left( \frac{1}{4}-q_{1}^{2}\right) \left(
\frac{1}{4}-q_{2}^{2}\right) e^{-a(q_{1}-q_{2})^{2}}  \nonumber \\
\varphi _{A_{u}}(q_{1},q_{2}) &=&\left( \frac{1}{4}-q_{1}^{2}\right) \left(
\frac{1}{4}-q_{2}^{2}\right) (q_{1}+q_{2})e^{-a(q_{1}-q_{2})^{2}}  \nonumber
\\
\varphi _{B_{u}}(q_{1},q_{2}) &=&\left( \frac{1}{4}-q_{1}^{2}\right) \left(
\frac{1}{4}-q_{2}^{2}\right) (q_{1}-q_{2})e^{-a(q_{1}-q_{2})^{2}}  \nonumber
\\
\varphi _{B_{g}}(q_{1},q_{2}) &=&\left( \frac{1}{4}-q_{1}^{2}\right) \left(
\frac{1}{4}-q_{2}^{2}\right) \left( q_{1}^{2}-q_{2}^{2}\right)
e^{-a(q_{1}-q_{2})^{2}}.  \label{eq:fi_var_bet_1}
\end{eqnarray}

\section{Results\label{sec:results}}

\begin{figure}[H]
\begin{center}
\includegraphics[width=9cm]{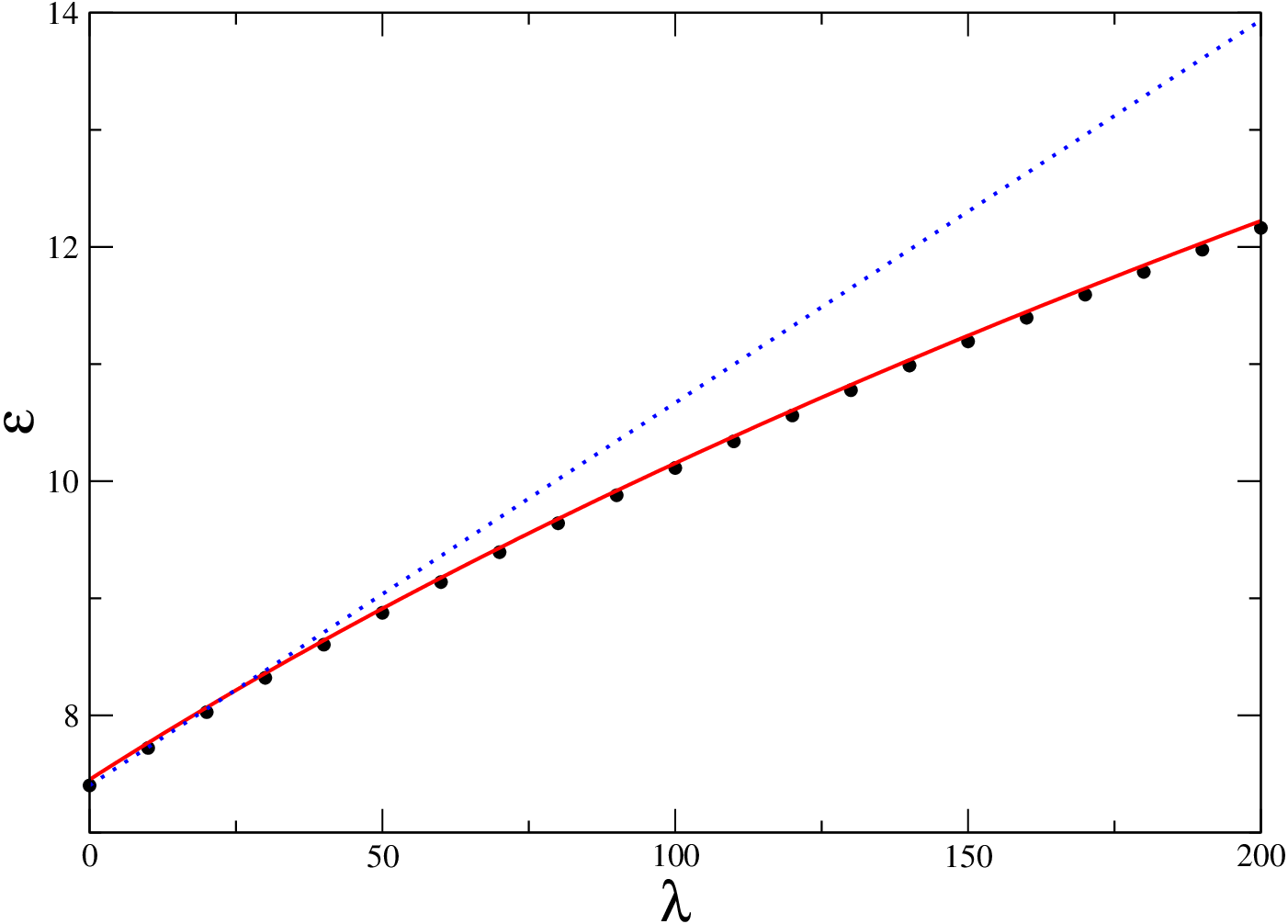}
\end{center}
\caption{Perturbation theory (dotted line), LSF (points) and variational
(solid line) calculation of the ground--state energy $\epsilon(\lambda)$ for
$\beta=1/2$.}
\label{fig:bet05a}
\end{figure}

\begin{figure}[H]
\begin{center}
\includegraphics[width=9cm]{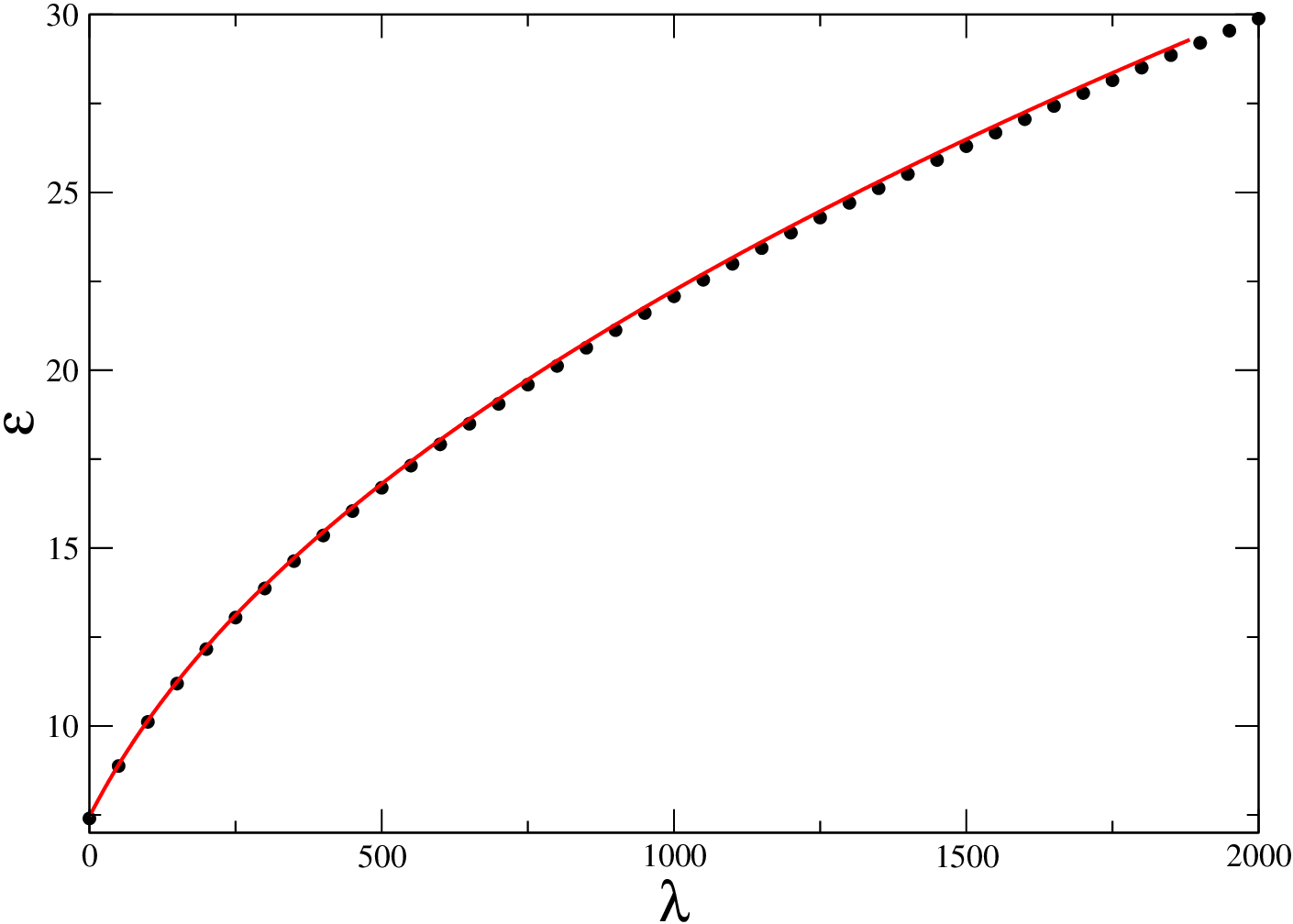}
\end{center}
\caption{Variational (line) and LSF (points) calculation of the
ground--state $\epsilon(\lambda)$ for $\beta=1/2$.}
\label{fig:bet05b}
\end{figure}

\begin{figure}[H]
\begin{center}
\includegraphics[width=9cm]{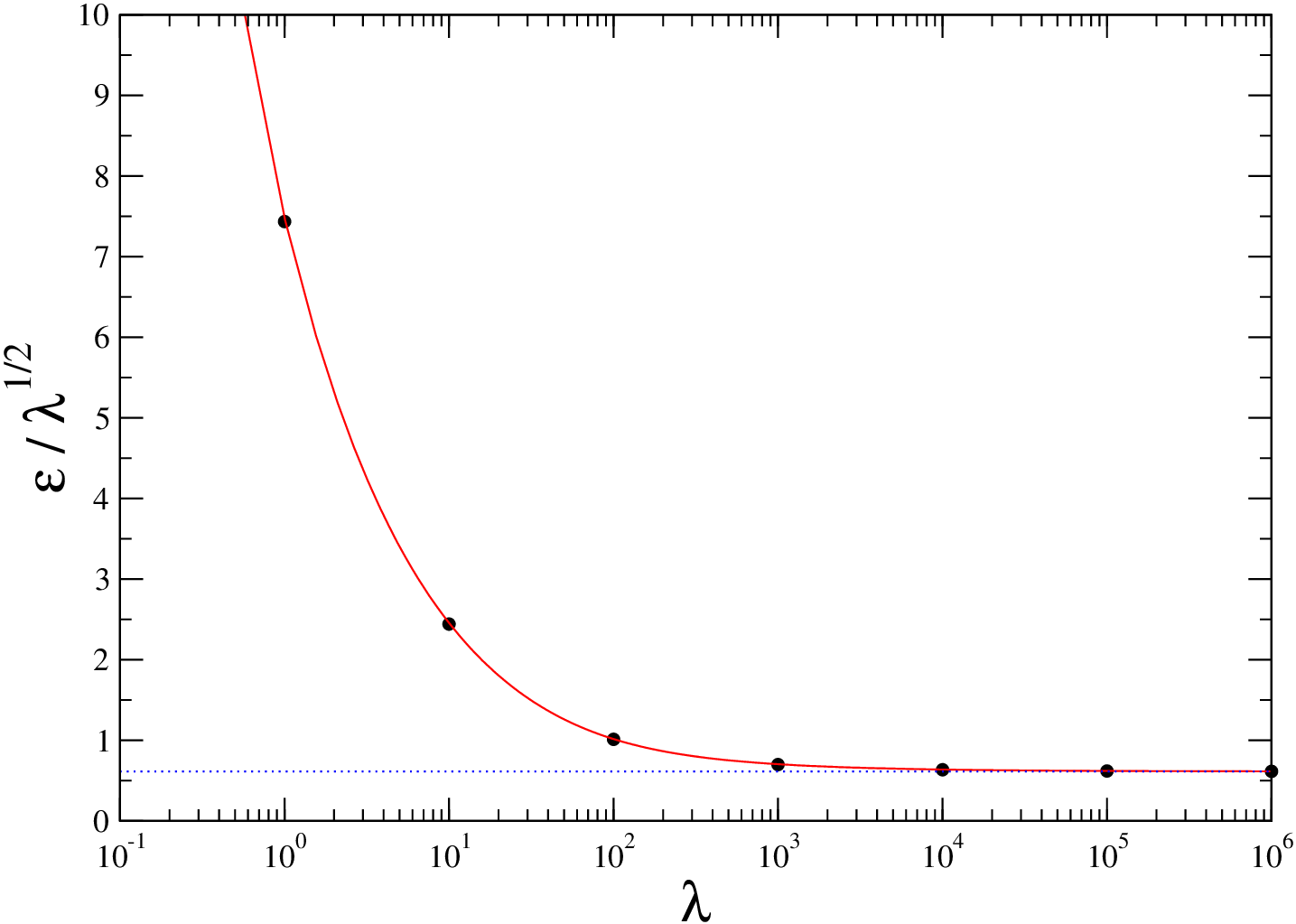}
\end{center}
\caption{Variational (line) and LSF (points) calculation of the
ground--state $\epsilon(\lambda)/\protect\sqrt{\lambda}$ for $\beta=1/2$.
The horizontal line marks the limit $\protect\sqrt{3/8}$.}
\label{fig:bet05_limit}
\end{figure}

\begin{figure}[H]
\begin{center}
\includegraphics[width=9cm]{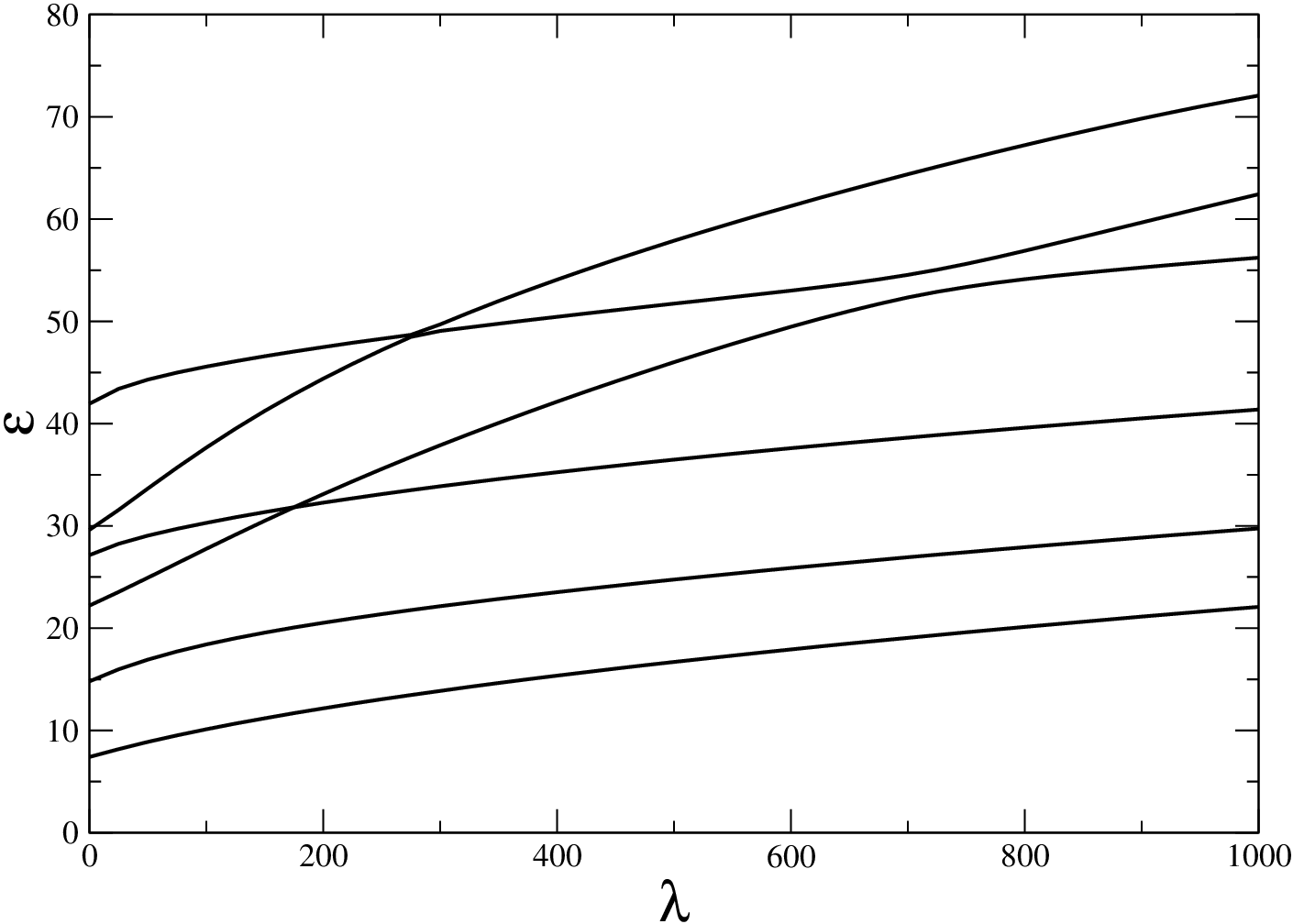}
\end{center}
\caption{First six eigenvalues for $\beta=1/2$ calculated by means of the
LSF method. }
\label{fig:bet05s}
\end{figure}

Fig.~\ref{fig:bet05a} shows the ground--state energy for $\beta =1/2$
calculated by means of perturbation theory, the LSF method and the
variational function (\ref{eq:fi_var_bet<>1}) for small and moderate values
of $\lambda $. Fig.~\ref{fig:bet05b} shows the results of the latter two
approaches for a wider range of values of $\lambda $. We appreciate the
accuracy of the energy provided by the simple variational function (\ref
{eq:fi_var_bet<>1}) for all values of $\lambda $. The reader will find all
the necessary details about the LSF collocation method elsewhere\cite{AF09}.
Here we just mention that a grid with $N=60$ was sufficient for the
calculations carried out in this paper. Fig.~\ref{fig:bet05_limit} shows
that $\epsilon (\lambda )/\sqrt{\lambda }$ calculated by the same two
methods for $\beta =1/2$ approaches $\sqrt{3/8}$ as suggested by equation (%
\ref{eq:e_lim}). Fig.~\ref{fig:bet05s} shows the first six eigenvalues $%
\epsilon (\lambda )$ for $\beta =1/2$ calculated by means of the LSF
collocation method. The level order to the left of the crossings is $%
\epsilon _{1}(A_{g})<\epsilon _{1}(A_{u})<\epsilon _{2}(A_{u})<\epsilon
_{2}(A_{g})<\epsilon _{3}(A_{g})<\epsilon _{3}(A_{u})$. Notice the crossings
between states of different symmetry and the avoided crossing between the
states $2A_{u}$ and $3A_{u}$.

\begin{figure}[H]
\begin{center}
\includegraphics[width=9cm]{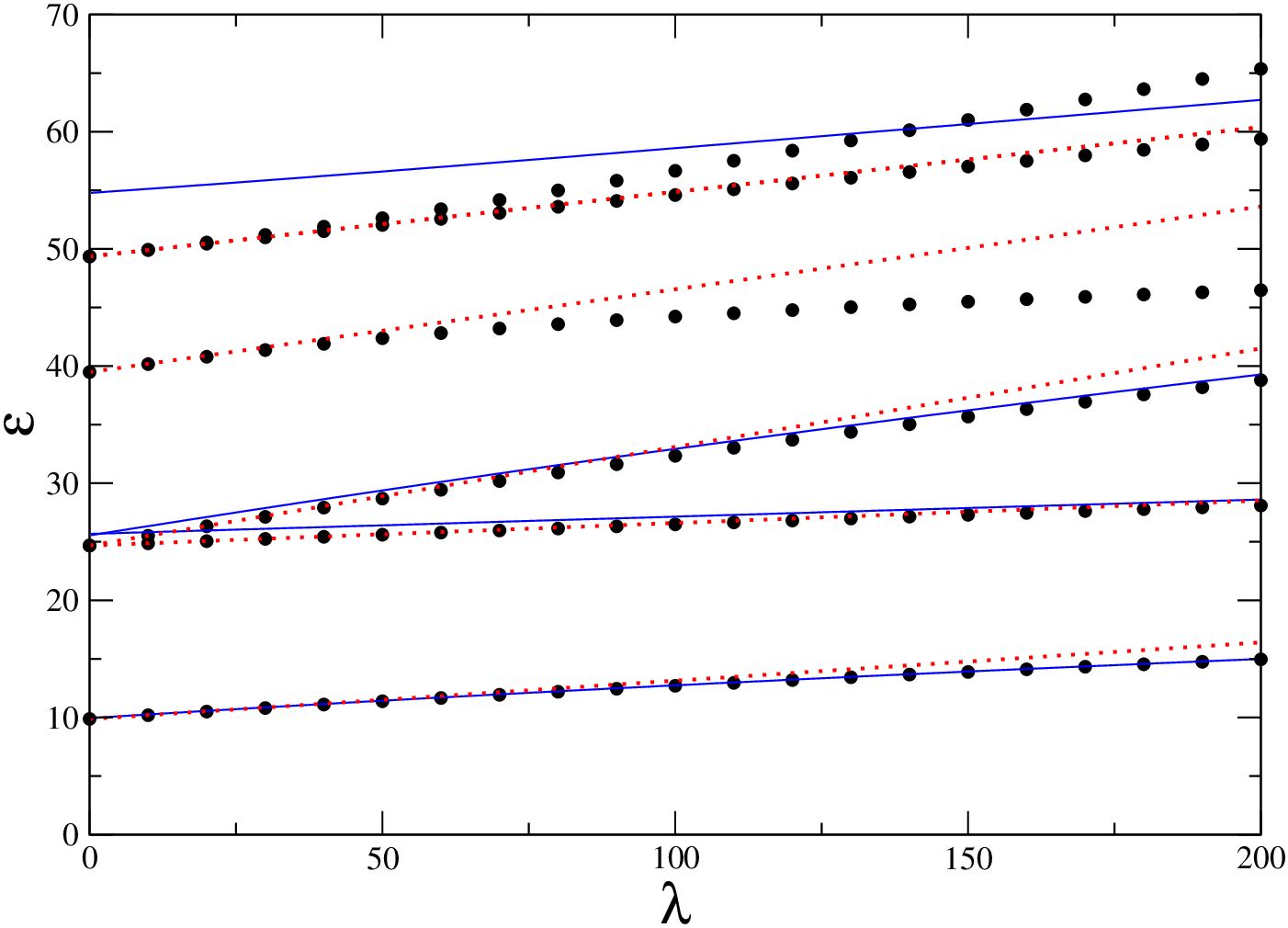}
\end{center}
\caption{First six eigenvalues for $\beta=1$. Circles, dotted line and solid
line correspond to LSF collocation approach, perturbation theory and
variation method, respectively. The level order is $%
A_g<A_u<B_u<2A_g<3A_g<B_g $.}
\label{fig:bet1a}
\end{figure}

\begin{figure}[H]
\begin{center}
\includegraphics[width=9cm]{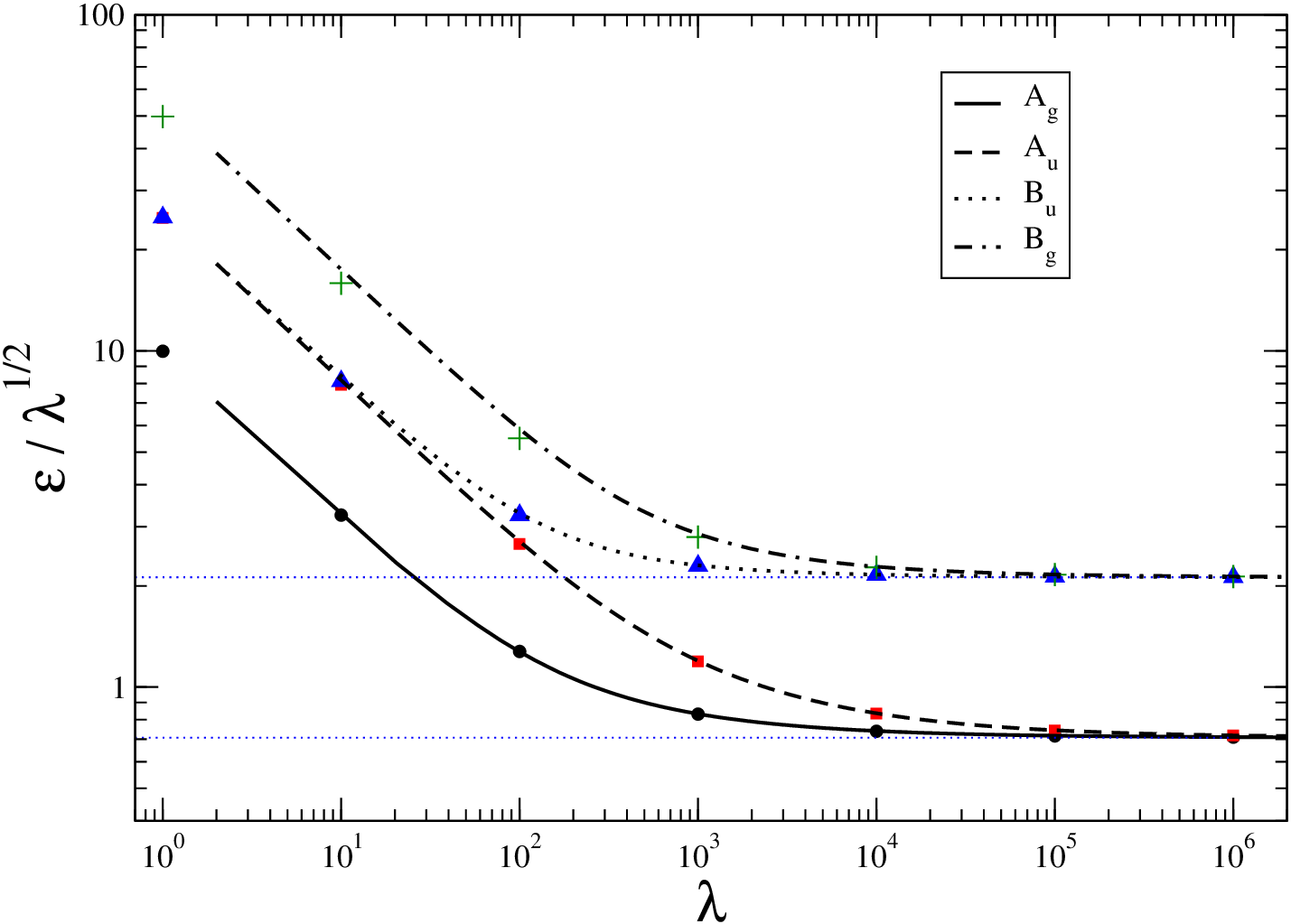}
\end{center}
\caption{Variational (solid line) and LSF (symbols) calculation of $%
\epsilon(\lambda)/\protect\sqrt{\lambda}$ for $\beta=1$. The horizontal
lines mark the limits $1/\protect\sqrt{2}$ and $3/\protect\sqrt{2}$.}
\label{fig:limit}
\end{figure}

Fig.~\ref{fig:bet1a} shows the first six eigenvalues for $\beta =1$
calculated by means of perturbation theory, the LSF method and the
variational functions (\ref{eq:fi_var_bet_1}) for small values of $\lambda $%
. The energy order is $\epsilon _{1}(A_{g})<\epsilon _{1}(A_{u})<\epsilon
_{1}(B_{u})<\epsilon _{2}(A_{g})<\epsilon _{3}(A_{g})<\epsilon _{1}(B_{g})$
and we appreciate the splitting of the energy levels $\epsilon _{3}(A_{g})$
and $\epsilon _{1}(B_{g})$ that does not take place at first order of
perturbation theory as discussed in Sec.~\ref{sec:small_L}. Finally, Fig.~%
\ref{fig:limit} shows $\epsilon (\lambda )/\sqrt{\lambda }$ for sufficiently
large values of $\lambda $. We appreciate that the four simple variational
functions (\ref{eq:fi_var_bet_1}) are remarkably accurate and that $\epsilon
(\lambda )/\sqrt{\lambda }\rightarrow 1/\sqrt{2}$ for the first two states
of symmetry $A_{g}$ and $A_{u}$ and $\epsilon (\lambda )/\sqrt{\lambda }%
\rightarrow 3/\sqrt{2}$ for the next two ones of symmetry $B_{u}$ and $B_{g}$%
. These results are consistent with equation (\ref{eq:Psi_lim_bet=1}) that
suggests that the energies of the states with symmetry $A$ and $B$ approach $%
\sqrt{2}(2v+1/2)$ and $\sqrt{2}(2v+3/2)$, respectively. In fact, Fig.~\ref
{fig:limit} shows four particular examples with $v=0$.

\section{Conclusions\label{sec:conclusions}}

The model discussed in this paper is different from those considered before%
\cite{AK40,A41,A42,C43,AK45,D52,BS55,D66,V68,V73,CF76,AM77,R78,ALZ80,AILZ81,
BR81,FC81a,FC81b,AGZL83,CM83,ML83,A97,V98} because in the present case the
linear force is due to the interaction between two particles. Although the
interaction potential depends on the distance between the particles the
problem is not separable and should be treated as a two--dimensional
eigenvalue equation. It is almost separable for a sufficiently small box
because the interaction potential is negligible in such limit and also for a
sufficiently large box where the boundary conditions have no effect. It is
convenient to take into account the symmetry of the problem and classify the
states in terms of the irreducible representations because it facilitates
the discussion of the connection between both regimes.

The model may be suitable to investigate the effect of pressure on the
vibrational spectrum of a diatomic molecule and in principle one can
calculate the spectral lines by means of the Rayleigh--Ritz or the LSF
collocation method\cite{AF09}.

The simple variational functions developed some time ago \cite{AFC84} and
adapted to present problem in Sec.~\ref{sec:large_L} provide remarkably
accurate energies for all values of the box size and are, for that reason,
most useful to show the connection between both regimes and to verify the
accuracy of more elaborate numerical calculations.

\end{document}